\documentclass[amsmath,amssymb,twocolumn,notitlepage,nofootinbib,superscriptaddress,floatfix,eprint]{revtex4-2}
\usepackage[utf8]{inputenc}
\usepackage{graphicx}
\usepackage{braket}
\usepackage{multirow}
\usepackage{booktabs}
\usepackage{hyperref}
\usepackage{xcolor}
\usepackage{float}
\usepackage{subcaption}
\usepackage{caption}
\usepackage{bm}
\usepackage{soul}
\usepackage{mathtools}
\usepackage{makecell} 
\usepackage{microtype}
\usepackage{url}

\begin{document}
%\linenumbers

\title{Deployment of Entanglement-Based QKD in Financial Infrastructure}

\author{Mirela Selimovi\'{c}}
\email{mirela.selimovic@univie.ac.at}
\affiliation{zerothird GmbH, Clemens-Holzmeister-Straße 6/6, 1100 Vienna, Austria}
\affiliation{University of Vienna, Faculty of Physics, Vienna Center for Quantum
Science and Technology (VCQ), Boltzmanngasse 5, Vienna 1090, Austria}
\affiliation{Vienna Doctoral School in Physics (VDSP), Boltzmanngasse 5, 1090 Vienna, Austria}

\author{Roman Solar}
\affiliation{qtlabs GmbH, Clemens-Holzmeister-Straße 6/6, 1100 Vienna, Austria}
\affiliation{Atominstitut, Technische Universität Wien, Stadionallee 2, 1020 Wien, Austria.}

\author{Jonathan Gruner}
\author{Sebastian Mair}
\author{Mario Wenzl}
\author{Thomas Heine}
\affiliation{zerothird GmbH, Clemens-Holzmeister-Straße 6/6, 1100 Vienna, Austria}

\author{Matej Pivoluska}
\affiliation{qtlabs GmbH, Clemens-Holzmeister-Straße 6/6, 1100 Vienna, Austria}

\author{Rupert Ursin}
\affiliation{zerothird GmbH, Clemens-Holzmeister-Straße 6/6, 1100 Vienna, Austria}
\affiliation{qtlabs GmbH, Clemens-Holzmeister-Straße 6/6, 1100 Vienna, Austria}

\author{Sebastian Philipp Neumann}
\email{sebastian.neumann@zerothird.com}
\affiliation{zerothird GmbH, Clemens-Holzmeister-Straße 6/6, 1100 Vienna, Austria}

\begin{abstract}
We demonstrate the feasibility of entanglement-based quantum key distribution (eQKD) in high-security financial infrastructure over a 22\,km fiber link with 8\,dB loss between two data centers using polarization entanglement. The fully automated system continuously generated secure keys for four months at an average rate of 63.8\,kb/s, which were stored into a key management system and consumed to establish a VPN tunnel. The setup achieved 93.7\,$\%$ total up-time, with no downtime caused by the quantum optical components. Active polarization control kept the quantum bit error rate below 2\,$\%$ for 97.4\,$\%$ of the time and timing synchronization based on the entangled photon pairs' intrinsic temporal correlations achieved sub-300\,ps precision. Our standalone system requires neither polarized guide lasers nor external high-precision time references. These results show practical integration of eQKD into operational financial infrastructure.
\end{abstract}

 \maketitle

\section{Introduction}
%\sn{ würd den Fehler weglassen. Außerdem wird normalerweise ein Fehler auf 2 signifikante Stellen angegeben und dann die Zahl auf dieselben Stellen gerundet, also in dem Fall: 63.8$\pm$1.0. Diese Regel bitte auch im Folgenden anwenden, wo ein Fehler angegeben wird}
The security of modern communication networks relies predominantly on classical cryptographic schemes, which are vulnerable to the computational capabilities of emerging quantum computers \cite{grover_fast_nodate, shor_polynomial-time_1997, montanaro_quantum_2016, boixo_characterizing_2018, harrow_quantum_2009}. Quantum key distribution (QKD) offers a secure method to establish cryptographic keys by leveraging the principles of quantum mechanics \cite{xu_secure_2020}, providing long-term security independent of computational assumptions \cite{lo_secure_2014}. Among the various QKD protocols, entanglement-based schemes exploit both the nonlocal correlations of entanglement and the no-cloning theorem to generate shared cryptographic keys \cite{ekert_quantum_1991, bennett_quantum_1992}.
Despite significant progress in laboratory demonstrations and field trials \cite{pelet_operational_2023, craddock_automated_2024, kucera_demonstration_2024, sena_high-fidelity_2025}, the deployment of entanglement-based QKD (eQKD) in operational infrastructure remains limited. In particular, integrating such systems into existing metropolitan fiber networks of critical institutions introduces challenges, including channel losses, environmental perturbations, the need for reliable long-term operation, and overcoming organizational challenges. To address these challenges, we designed and deployed an eQKD system based on polarization entanglement that connects the operational data centers of a financial institution \cite{erste2026quantum}. This work demonstrates error-corrected and privacy-amplified key handover to a key management system (KMS) over real-world fiber links, incorporates active polarization control to maintain entanglement fidelity, and can by design support multi-user scalability through a wavelength division multiplexing scheme \cite{wengerowsky_entanglement-based_2018, joshi_trusted_2020}. This deployment establishes a reference for the maturity of eQKD in operational environments and provides a foundation for the development of scalable eQKD networks.

This work is structured as follows: \ref{sec:qkd} reviews quantum key distribution, \ref{sec:systemarch} details the system architecture including link design, classical data processing and secret key distillation, \ref{sec:performance} presents performance evaluation, and \ref{sec:conclusion} discusses the results.

\section{Quantum key distribution}
\label{sec:qkd}
\begin{figure*}[htp]
    \centering
    \includegraphics[width=0.90\textwidth]{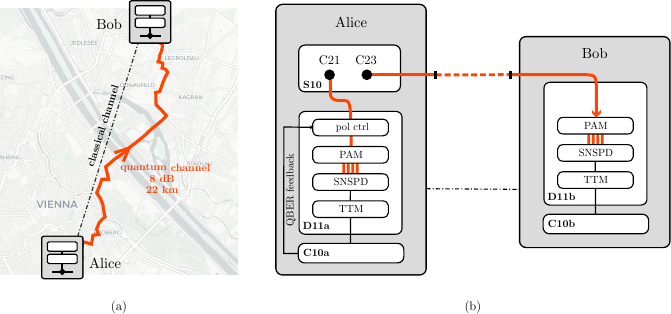}
    %\hspace{5mm}
    %\begin{minipage}[b]{0.33\textwidth}
    %    \centering
    %    \includegraphics[width=\textwidth]{figures/p2p-link-FIG1-A-v2.pdf}
        
    %    \vspace{5mm}
    %    (a)
    %\end{minipage}
    %\hfill
    %\begin{minipage}[b]{0.55\textwidth}
    %    \centering
    %    \includegraphics[width=\textwidth]{figures/p2p-link-FIG1-B.pdf}
    
    %    \vspace{1mm}
    %    (b)
    %\end{minipage}
    \hspace{5mm}
    \caption{\textbf{Schematic of entanglement-based QKD operation over a deployed fiber link between two Viennese data centers of a financial institution.} (a) Map showing the geographical locations of the data centers, Alice and Bob, connected by a fiber link with an overall loss of 8\,dB. (The red line is for illustration only and does not represent the actual fiber deployment.) (b) Point-to-point connection: Alice is co-located with the entangled photon-pair source (S10), while Bob is connected via the deployed fiber link (quantum channel). The detection modules (D11a and D11b) at both communication partners are identical, except that Alice's module includes a polarization control module (pol ctrl) before the photon measurement stage. The fully passive polarization analysis module (PAM) projects each incoming photon onto either the rectilinear or diagonal polarization basis and directs it to the corresponding channel of a superconducting nanowire single-photon detector (SNSPD). Photon detection times are recorded by a time tagger module (TTM) for each of the four channels. Two servers (C10a and C10b) distill secure keys from the registered detection events through a classical communication channel. Polarization compensation becomes active when the QBER exceeds 2\%.}
    \label{fig:conceptual}
\end{figure*}

Currently used asymmetric cryptographic algorithms, such as RSA, derive their security from the assumed computational difficulty of, for instance, large‑integer factorization \cite{rivest_method_1978}. In contrast, QKD enables two distant parties to establish shared secret keys whose security is rooted in fundamental principles of quantum mechanics rather than computational hardness assumptions. In recent years, QKD has established its proof-of-principle feasibility as a technology for secure communication infrastructures. Among the various QKD approaches, such as continuous-variable \cite{diamanti_distributing_2015}, measurement-device independent \cite{liu_experimental_2019, berrevoets_deployed_2022} or twin-field \cite{wang_twin-field_2022, liu_experimental_2023} QKD, entanglement-based schemes offer a distinct paradigm in which correlated measurement outcomes originate from entangled quantum states, enabling security guaranties that are linked to nonlocal quantum correlations \cite{waks_security_2002}. By exploiting the disturbance induced by quantum measurements, legitimate eQKD users can detect the presence of an eavesdropper and bound the potentially leaked information during key exchange. Note that in contrast to prepare-and-measure protocols \cite{bennett1984quantum, shor2000simple}, eQKD does not require trust in the state preparation. 

The deployment of eQKD over optical fiber infrastructure introduces a number of practical challenges that directly affect system performance and stability. Optical attenuation limits the achievable distribution distance due to photon loss, chromatic dispersion can degrade temporal correlations required for efficient key generation, and active time synchronization is necessary to correct for clock drift in remote detection systems. Additionally, in systems based on polarization entanglement, environmental perturbations of the transmission fiber represent a critical factor. Mechanical stress and temperature fluctuations can induce birefringence variations and therefore polarization rotations that must be actively compensated to preserve the fidelity of the entangled state.
In this work, we present a field-deployed eQKD setup operating stably across real-world fiber links connecting two data centers of a financial institution \cite{erste2026quantum}, using polarization-entangled quantum states. The deployed system successfully produced post-processed secure keys according to the BBM92 QKD protocol \cite{bennett_quantum_1992} at rates of 63.78$\pm$1.02\,kb/s on average with an up-time of 93.7\,$\%$ over 2800\,h continuous operation. The generated keys were used to establish a VPN tunnel between the data centers.

\begin{figure*}[!t]
    \centering
    \includegraphics[width=0.95\textwidth]{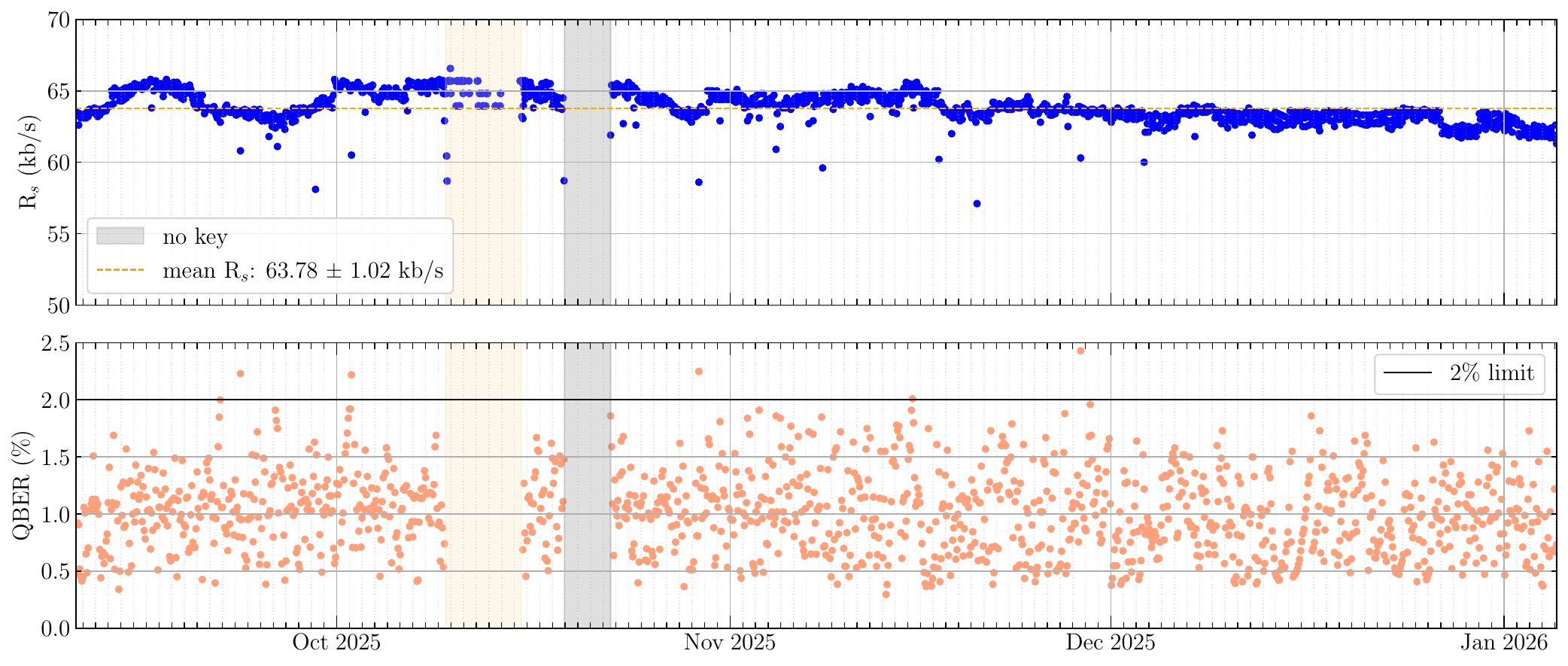}
    \caption{\textbf{Operation of entanglement-based QKD system in financial infrastructure over 2800\,h (4\,months)} The system had an up-time of about 93.7\,\%. Top: Secure key rate over the measurement period 63.78$\pm$ 1.02\,kb/s on average. The gray shaded region indicates a three-day interval without key creation due to a test power outage. The yellow shaded region marks an eight-day period of unstable operation following changes in the network configuration in the data center. These changes resulted in a temporarily reduced up-time of 44\,\%; also, QBER values were not recorded during this time. The secure key rate R$_s$ decreases below the long-term average of the operational period; this is attributable to aging of the source's pump laser diode and can be compensated for by increasing the diode current. Bottom: QBER over a period of 4 months of continuous operation. Random polarization rotation in the fiber occurs mainly due to temperature drifts. These are compensated actively with a polarization controller that is initiated whenever QBER reaches values $\ge$ 2\,\%.}
    \label{fig:key-and-qber}
\end{figure*}

\section{System architecture}
\label{sec:systemarch}
The architecture of the deployed eQKD system was designed to accommodate the constraints of real-world fiber links while maximizing stability and key generation performance. The main effects that limit performance in fiber links are polarization rotation, chromatic dispersion, and timing synchronization.

Alice and Bob refer to two data centers of the financial institution. The system comprises a source of entangled photon pairs (S10) and two detector units (D11A and D11B, respectively). S10 and D11A were located at Alice, D11B at Bob. The two data centers are directly connected through a dark fiber (cf. FIG. \ref{fig:conceptual} (a)) carrying the photonic quantum states. The source of entangled photon pairs generates polarization-entangled photon pairs that lead to correlated outcomes in accordance with the maximally entangled Bell state
\begin{equation}
    \ket{\Phi^+}_{AB} = \frac{\ket{H}_A \otimes\ket{H}_B + \ket{V}_A \otimes \ket{V}_B}{\sqrt{2}}.
    \label{eq:phi+}
\end{equation}
One photon is detected at Alice, while the other is transmitted to Bob over the quantum channel exhibiting approximately 8\,dB of link loss (cf. FIG. \ref{fig:conceptual} (b)). Note that the source does not necessarily have to be co-located with Alice. The best performance in terms of key generation is expected with symmetric loss setups \cite{neumann_model_2021}. 

\subsection{Link design}
\label{subsec:setup}
The entanglement source generates polarization-entangled photon pairs in the telecom C-band via type-0 spontaneous parametric down-conversion (SPDC)\cite{ortega_spatial_2023}, centered at 1559.77\,nm. Entangled photon pairs are produced by pumping a passively stable interferometric module with an industry-grade laser operating at 779.89\,nm. The interferometric module, with dimensions of 100\,mm $\times$ 100\,mm, is a highly integrated, alignment-free optical assembly that ensures long-term operational stability and robustness. The optimized optical design provides a spectral pair generation rate of 6.8\,Mcps/mW/nm, visibility $\geq$ 99.4\,$\%$ and heralding efficiency $\geq$ 65\,$\%$. These metrics remained stable throughout the entire operating period, reflecting the robustness of the passively stable design. By spectral filtering using standard dense wavelength-division multiplexing equipment, we select two 100\,GHz broad channels from the generated SPDC spectrum; channel C21 (according to the ITU grid \cite{ITU_G6941_2020}) is measured locally with Alice, while the conjugate channel C23 is transmitted to Bob via the fiber link. The selection of conjugate wavelength channels symmetric to the center wavelength follows directly from energy conservation in SPDC, which produces frequency-correlated photon pairs \cite{zhang_spontaneous_2021}.

Alice and Bob are equipped with identical detection modules, except for the polarization control module, which is included in Alice's setup only. Incoming photons are analyzed using a polarization analysis module (PAM)that randomly projects each photon onto the rectilinear (H/V) or diagonal (D/A) polarization basis. The splitting ratio for Alice (Bob) is 49.6\,$\%$ (49.9\,$\%$) in the rectilinear and 50.4\,$\%$ (50.1\,$\%$) in the diagonal basis. This passive basis choice ensures mutually unbiased basis measurements without the use of active modulation. Each detection module comprises four superconducting nanowire single-photon detectors (SNSPD) with a detection efficiency of $\geq$ 70\,$\%$ to measure in two polarization bases. The detection unit including electronics and cryogenics modules are 19-inch, data center-compatible rack frames with a height of 14 rack units.

Alice and Bob record photon detection events as discrete timestamp sequences $\{t_i^{(A)}\}$ and $\{t_j^{(B)}\}$ using independent, unsynchronized time-tagging units. A relative clock offset  $\Delta(t)$ accumulates over time. Coincident events are identified by evaluating pairwise time differences $\kappa_{ij} = t_j^{(B)} - t_i^{(A)}$ and constructing a delay histogram 
\begin{equation}
    H(\kappa) = \sum_{i,j}{\delta(\kappa - (t_j^{(B)} - t_i^{(A)}))},
\end{equation}
which is equivalent to the cross-correlation of the two timestamp sequences. Coincident photon pairs give rise to a distinct coincidence peak in $H(\kappa)$ at $\kappa_0-\Delta(t)$, where $\kappa_0$ denotes the fixed system delay due to optical or electronic path differences.

Even for perfect timing synchronization, the width of this coincidence peak and therefore the precision of correlation depend on detector jitter and chromatic dispersion in the fiber. The SNSPDs feature low timing jitter in the order of 35\,ps. This leaves chromatic dispersion (CD) as the main contribution to temporal uncertainty. CD of 18.55\,ps/nm/km in the 22\,km deployed optical fiber results in a total temporal spread of 293.46\,ps. The combination of high pair generation, narrow spectral filtering of the photons and high-timing precision detection by the SNSPDs rendered CD compensation unnecessary. The pair generation rate of the source provides sufficient photon statistics to achieve a signal-to-noise ratio to resolve the coincidence peaks despite this temporal broadening. As CD compensation would add additional loss, in this particular case, it is beneficial to omit a dedicated compensation mechanism.

An important metric used throughout is the Quantum Bit Error Rate (QBER), defined as the fraction of mismatched bits between the two parties’ so-called sifted keys. It serves as an indicator of errors in the quantum channel, arising from noise, imperfections, or potential eavesdropping, and is used to assess the required amount of error correction and privacy amplification in classical data processing.

Once a common time base is established, the QBER is calculated using 10\,$\%$ of the coincident events. This QBER information is used only for polarization control and performance monitoring, not for key creation. For polarization-entangled photon pairs transmitted over optical fiber links, random polarization drifts increase QBER. Therefore, drifts are actively compensated using an in-fiber three-axis polarization controller module. 
Let \(U_A\) and \(U_B\) denote the local unitary polarization rotations induced by Alice's and Bob's optical fiber links, respectively. The compensation module, acting only on Alice's side, applies the unitary
\begin{equation}
    U_C = (U_B^T)^\dagger U_A^{-1}.
\end{equation}
Since
\begin{equation}
    (U_A\otimes U_B)\ket{\Phi^+}_{AB} = (U_AU_B^T\otimes I)\ket{\Phi^+}_{AB}
\end{equation}
is a well-known identity for the maximally entangled state \(\ket{\Phi^+}_{AB}\), applying \(U_C\) as defined in Eq.~(3) on Alice's side is sufficient to compensate the combined polarization drift. Note that \(U_C\) is not known a priori, and the compensation is performed iteratively by adjusting the three axes of the polarization controller module while using the QBER as feedback to minimize it.
Once this is achieved, the sifting process is started, i.e., events corresponding to mismatched bases are discarded, and those measured in the same basis are assigned to binary values according to a predefined rule (e.g. H/D\,$\to\,0$ and V/A\,$\to\,1$), resulting in the so-called sifted key.

\subsection{Classical Data Processing and Secret Key Distillation}
\label{subsec:postprocessing}

 %Classical data processing covers all classical computer and communication operations required to distill a secure key from the sifted key.
 The implemented system realizes the complete classical processing chain required for key distillation, including parameter estimation (PE), error correction (EC), error verification (EV), privacy amplification (PA), and classical message authentication (MA). 
 %At the same time, the present deployment does not aim to instantiate every auxiliary assumption of the composable security model in its strongest information-theoretic form.  
 The current implementation uses computationally secure mechanisms for MA and EV tags, and the random choices required by post-processing are not supplied by a dedicated, characterized quantum random number generator (QRNG).
 We therefore use the finite-key framework of Tomamichel and Leverrier \cite{tomamichel_2017} as a reference model for the secret-key-length estimates. 
 We also note that the direct use of this proof relies on additional assumptions on the measurement implementation, in particular equal detection efficiencies in the two bases and an exact 50/50 beam splitter. 
 These assumptions are not fundamental and can be lifted by incorporating the techniques presented in Wang et al. \cite{Wang2025f}. A fully composable implementation with information-theoretic MA, universal-hashing-based EV, characterized QRNG, and a complete treatment of these measurement-device assumptions is therefore left as a natural extension of the present system.

In the implemented system, for each sifted key block, a randomly selected fraction of the sifted key is revealed for PE and excluded from the raw key block. Specifically, we use roughly 10\,$\%$ of the sifted bits for PE, such that the remaining raw key block has length \(n=10^5\). QBER is computed from the revealed bits; blocks for which it exceeds the selected acceptance threshold of 4\,$\%$ are discarded.

After PE, Alice's and Bob's raw keys are not identical due to channel noise, device imperfections, and possible eavesdropper interference. The aim of EC is to publicly reveal auxiliary information about the raw keys that allows Bob to correct discrepancies between his block and Alice's. We use one-way non-interactive syndrome-based EC, where Alice shares information about her raw key, the syndrome of length $m$, with Bob and he proceeds to use it to correct his raw key so that it matches Alice's. 

Our EC scheme is based on low-density parity-check codes (LDPC) \cite{Gallager1962}. We have constructed a parity-check matrix (PC) optimized for reliable decoding up to 4\,\% QBER for a block size of $n=10^5$. The matrix is generated with the PEG algorithm \cite{hu_2002}. Alice multiplies the PC matrix by her raw key to obtain the syndrome of $m=28,000$ bits. She shares it with Bob on the authenticated classical channel. He uses the syndrome and his raw key block to infer Alice's raw key block. This is done using the sum-product algorithm (SPA) \cite{Hu_2001} with a fixed maximum number of iterations. If Bob's raw key is still not compatible with the syndrome after the maximum number of iterations, processing of the current block is stopped, and the raw key block is discarded.

To measure the performance of the chosen LDPC matrix, we define the EC efficiency taking into account the frame error rate (FER) \cite{mueller_2024},
\begin{equation}
    f_\mathrm{FER} = (1 - \mathrm{FER})f + \frac{\mathrm{FER}}{H_2(\text{QBER})},
\end{equation}
where $\mathrm{FER}$ is the fraction of EC blocks that are unsuccessfully decoded, $H_2(\text{QBER})$ is the binary Shannon entropy of the actual QBER, and the efficiency $f$ is defined as usual for syndrome-based EC: 
\begin{equation}
    f = \frac{m}{nH_2(\text{QBER})}.
\end{equation}
This definition penalizes unsuccessful frames by assigning them zero secret-key contribution. For the threshold QBER of 4\,\%, the decoding simulations give $f_\mathrm{FER} \approx 1.16$.

After EC, there remains a non-zero probability that Bob's decoder outputs an incorrect block that is nevertheless compatible with the received syndrome. This probability is bounded by the EV step. In the deployed implementation, Alice and Bob compute and compare a verification tag of their reconciled blocks using a computationally secure hash function over the authenticated classical channel. This step can be replaced by a tag generated from a 2-universal hash family. If the tags differ, the reconciled block is discarded.

The final step is PA, which produces the secret key. The goal is to remove any information potentially leaked through EC and EV, as well as any residual correlations with an eavesdropper. This is achieved by projecting the reconciled key onto a shorter key using a 2-universal hash function. 

Here we use the Circulant extractor~\cite{Foreman_2025} as the 2-universal hash family. It is implemented using the Number Theoretic Transform (NTT)~\cite{vanassche_2006}, allowing a raw key block of length $n$ to be hashed in time $\mathcal{O}(n\log n)$, while consuming only $n+1$ random bits per raw key block. The Circulant extractor requires the input length plus one to be prime; therefore, before PA we discard 10 bits from each reconciled block, resulting in an input length of $99{,}990$ bits.

The resulting secret keys are handed to the KMS where they are indexed and securely stored in a database and subsequently made available for use in cryptographic applications. The keys were consumed by a Cisco virtual router using the SKIP protocol to establish a VPN tunnel. Alternatively, an ETSI GS QKD 014-compliant interface was implemented for key consumption.

%In addition, the KMS supports key distribution via the SKIP protocol to a Cisco virtual router and provides an ETSI GS QKD 014-compliant interface for interoperability with external applications and network elements.

Within this finite-key reference model~\cite{tomamichel_2017}, projecting each reconciled block to $17,000$ bits gives a security parameter $\varepsilon \approx 10^{-20}$, while projecting to $25,000$ bits gives $\varepsilon \approx 10^{-7}$. 

\section{Performance evaluation}
\label{sec:performance}
\begin{figure}[!t]
    \centering
    \includegraphics[width=\linewidth]{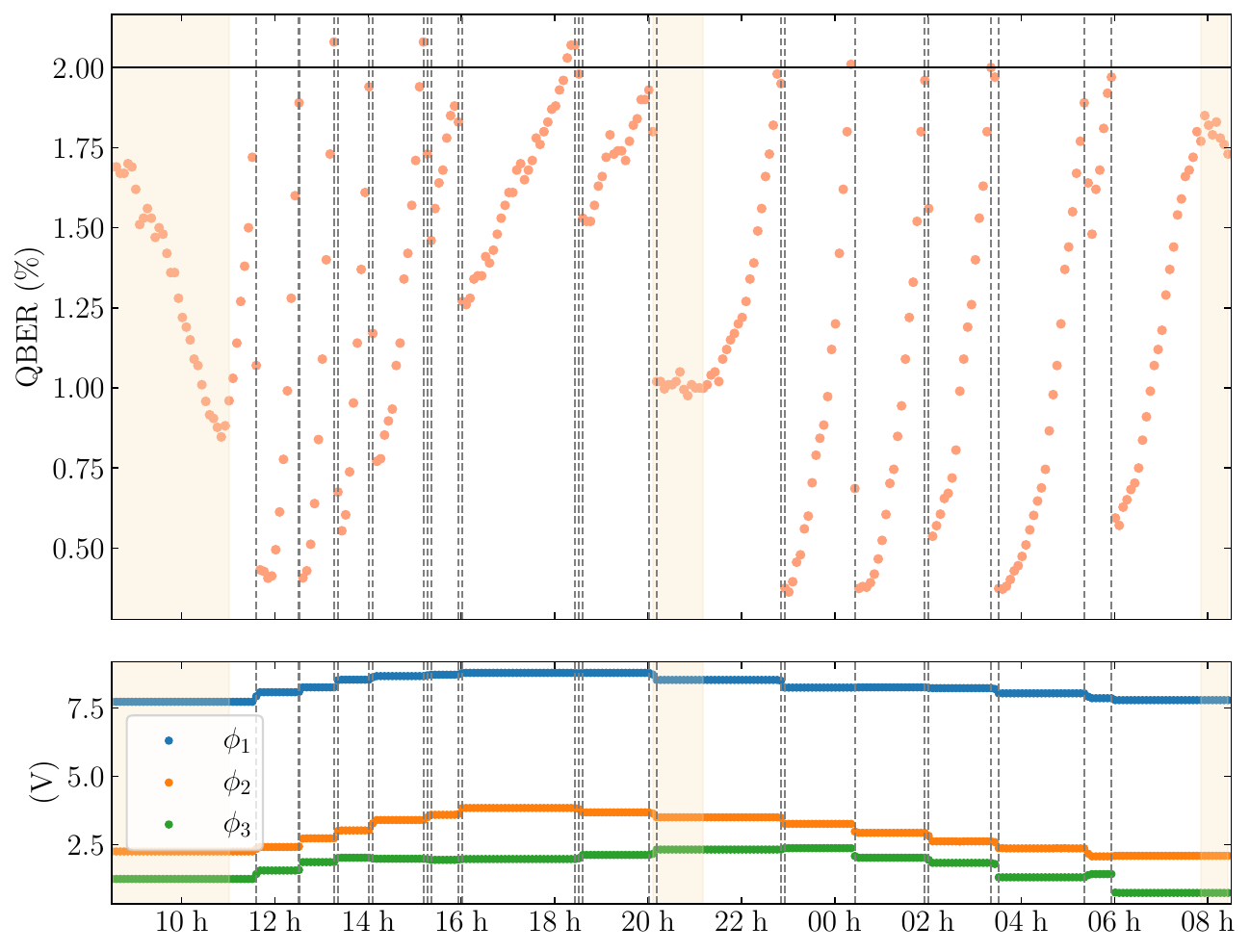}
    \caption{\textbf{Polarization rotation over a full day of operation in deployed fiber link.} The top panel shows the quantum bit error rate (QBER) as a function of time over one typical day of operation. Shaded regions indicate intervals where the QBER drifted in a favorable direction without intervention, while dashed lines mark instances where the polarization control algorithm is initiated to actively reduce the QBER. The algorithm is initiated once the QBER reaches $\geq$ 2\,\% (black horizontal line). The bottom panel displays the corresponding evolution of the three input voltages $\phi_1$, $\phi_2$ and $\phi_3$ of the polarization controller.}
    \label{fig:poldriftday}
\end{figure} 

For the long-term performance evaluation shown in FIG. \ref{fig:key-and-qber}, the PA output length was fixed to $52{,}552$ bits per post-processing block in order to characterize the throughput of the deployed post-processing pipeline and its integration with the KMS.

The system ran for 119 days with uninterrupted key generation for 108 days. During a period of 8 days, key generation was partially interrupted due to a test power outage. Here the system was up roughly half of the time generating keys that were transferred to the KMS. For additional three days, the system was completely down due to changes in the network configuration. In total, this makes an up-time of about 93.7\,\%, with none of the down-time being related to the quantum optical part of the setup. The interruptions due to software imperfections were rapidly resolved upon detection.

The theoretical maximum QBER to still extract key is 11\,$\%$ \cite{shor2000simple}; we chose an LDPC matrix for $\le$\,4\% as a compromise between correction efficiency and demands on polarization control. In our application, when QBER values reach 2\,\% or higher, the polarization control algorithm is activated and compensates in real time, using only QBER data as input, without the need of an external reference. FIG. \ref{fig:poldriftday} shows the QBER during a full day of operation as well as the corresponding activity of the three control axes of the polarization controller. Over the full operation time, the QBER was below 2\,\% for 97.4\,\% of the time, and higher QBER rates were compensated within an average of approximately 4\,minutes. 

\begin{figure}
    \centering
    \includegraphics[width=\linewidth]{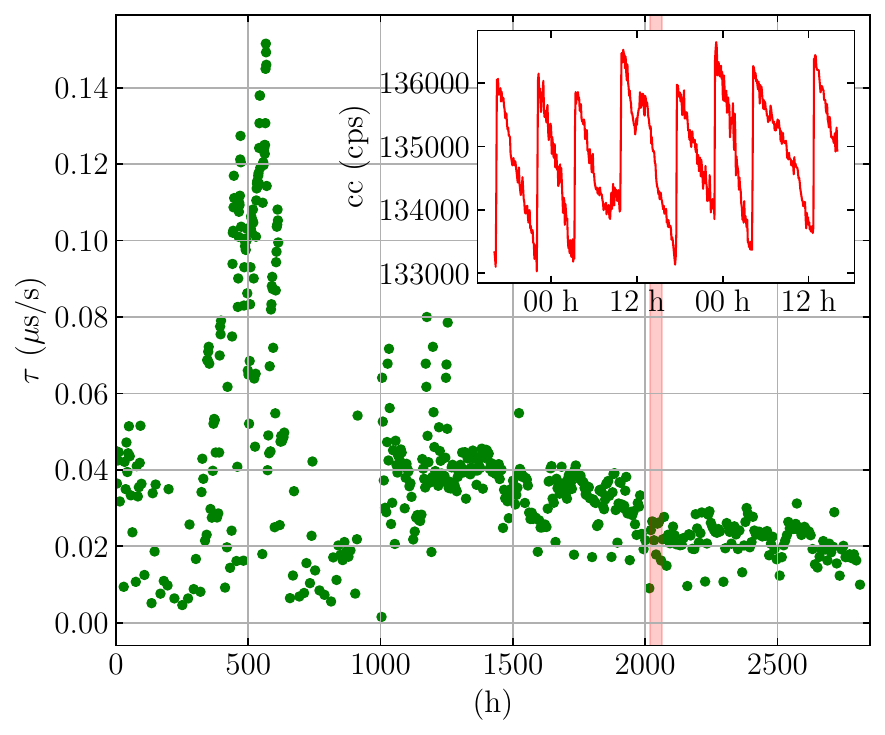}
    \caption{\textbf{Relative clock drift rate over four months of continuous operation.} The graph shows the relative time drift rate, $\tau$ (\textmu s/s), over the course of operation, resulting from the independent drift of the time-tagging clocks. The inset plot shows the coincidence count rate over two days of operation (red shaded region in main plot). Whenever the time drift reaches 500\,\textmu s, a re-synchronization event is initiated, resulting in a higher coincidence count rate and a sawtooth pattern.}
    \label{fig:clockdrift}
\end{figure}

Two free-running remote clocks accumulate small differences over time $\Delta(t)$ in their frequencies, leading to relative clock drifts.  Over time this effect offsets the time tag blocks of Alice and Bob with respect to each other, effectively making valid detection events in e.g. Alice's block lose their partner in Bob's block. FIG. \ref{fig:clockdrift} shows the clock drift rate $\tau$ over the four months of continuous operation with an inset plot that depicts the sawtooth pattern of the coincidence count rate between Alice and Bob. Once a relative time drift of 500\,\textmu s is reached, the system initiates a re-synchronization event. This produces a sawtooth pattern in the coincidence count rate and the secure key rate. The clock drift rate was higher in the first half of the operation period. This can be traced back to a lower ambient temperature of about 2$^{\circ}$C, as the time taggers' quartz clocks are temperature sensitive. Nevertheless, this effect could be easily compensated by computational methods only.

\section{Conclusion}
\label{sec:conclusion}
This study demonstrates the practical feasibility of eQKD in the high-security environment of a financial institution's data centers, achieving stable key generation over four months with low QBER and high system reliability. The hands-off high-performance source of polarization-entangled photon pairs was passively stable and showed no signs of degradation during operation time. The presented system operates without external timing references or auxiliary polarization signals, relying instead on intrinsic correlations of entangled photon pairs and real-time feedback based on QBER. This level of autonomy is essential for scalable deployment in realistic environments subject to continuous perturbations. The efficient implementation of LDPC-based reconciliation with an effective error-correction efficiency of \(f_\mathrm{FER}\approx 1.16\) at the \(4\%\) QBER threshold contributed to maintaining a practical secret-key yield. The generated keys were successfully integrated into a KMS and tested by establishing a VPN tunnel, illustrating the integration with existing networking infrastructure. The underlying entangled photon-pair source supports multi-user connections, which could expand such an eQKD scenario scalable to fully connected networks with up to 10 users by use of wavelength division multiplexing \cite{joshi_trusted_2020}. Future work will address the coexistence of quantum and classical communication channels within shared fiber infrastructure, further improving deployability. Advances in photonic integration, detector technology, and network orchestration are expected to significantly enhance key rates and scalability. Overall, our results demonstrate that entanglement-based QKD is no longer confined to experimental testbeds but is ready to play a practical role in securing critical communication infrastructure.

\section*{Acknowledgments}
We thank Erste Digital and A1 for providing access to the data center and  fiber infrastructure facilities required for the field deployment of the quantum key distribution system.
We are grateful to the technical staff at Erste Group and A1 for their assistance in installation and operational support of the optical network, especially to Martin Kukacka, Thomas Kraner, and Martin Ofner for their helpful support and collaboration during the project.

\section*{Author Contribution}
M.S. designed and conducted the data analysis study, interpreted the results, and wrote the manuscript. M.S., T.H. and S.N. developed the devices used in the proof-of-concept experiment and contributed technical expertise and experimental support. J.G., S.M. and M.W. implemented and developed the software for classical data processing. R.S. and M.P. supported the implementation of software from the quantum communication security perspective and supported the implementation of related components. R.S., M.P. and S.N. also contributed to writing the manuscript. T.H., R.U. and S.N. supervised the project. All authors approved the final version.

\section*{Competing interests}
The authors declare no competing interests. M.S. is affiliated with University of Vienna and an employee of zerothird GmbH. Erste Digital and A1 provided access to operational data center facilities and fiber infrastructure for the field deployment. Neither had any role in the data analysis, interpretation of results, or manuscript preparation.

\bibliographystyle{naturemag}
\bibliography{p2p-lib}

\end{document}